\documentclass[aps,prd,twocolumn,amssymb,showpacs,eqsecnum]{revtex4}
\usepackage{amsmath}
\usepackage{pstricks,pst-node,pst-plot}
\renewcommand{\(}{\left(}
\renewcommand{\)}{\right)}
\renewcommand{\[}{\left[}
\renewcommand{\]}{\right]}
\begin{document}


\title{Classical and quantum radiation reaction for linear acceleration}

\author{ Atsushi Higuchi$^1$ and Giles~D.~R.~Martin$^2$}

\affiliation{Department of Mathematics, University of York,
Heslington, York YO10 5DD, UK\\ email: $^1$ah28@york.ac.uk,
$^2$gdrm100@york.ac.uk}

\date{January 5, 2005}

\begin{abstract}
We investigate the effect of radiation reaction on the motion of a
wave packet of a charged scalar particle linearly accelerated in
quantum electrodynamics. We give the details of the calculations
for the case where the particle is accelerated by a static
potential that were outlined in Phys.Rev. D 70 (2004) 081701(R)
and present similar results in the case of a time-dependent but
space-independent potential. In particular, we calculate the
expectation value of the position of the charged particle after
the acceleration, to first order in the fine structure constant in
the $\hbar \to 0$ limit, and find that the change in the
expectation value of the position (the position shift) due to
radiation reaction agrees exactly with the result obtained using
the Lorentz-Dirac force in classical electrodynamics for both
potentials. We also point out that the one-loop correction to the
potential may contribute to the position change in this limit.
\end{abstract}


\maketitle

\section{Introduction}

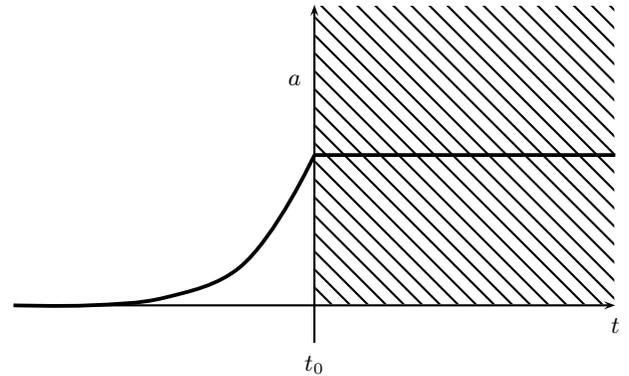
\begin{figure}
\begin{center}
\begin{pspicture}(-4,-1)(4,4)
 \psline{->}(-4,0)(4,0) \psline{->}(0,-0.5)(0,4)
 \pscurve[linewidth=0.5mm]{-}(-4,0)(-3,0)(-2,0.1)(-1,0.5)(0,2)
 \psline[linewidth=0.5mm]{-}(0,2)(4,2)
 \uput[l](0,3){$a$} \uput[d](4,0){$t$} \uput[d](0,-0.5){$t_0$}
 \psframe[linestyle=none, fillstyle=vlines](0,0)(4,4)
\end{pspicture}
\end{center}
\caption{Preacceleration of a charged particle} \label{preacceleration}
\end{figure}

A charged particle radiates when it is accelerated. This simple
discovery has been the basis of a multitude of technological
advances since. Most work, however, has concentrated on the
radiation itself. The effect of the radiation, which carries
energy and momentum, on the particle
itself is very small but still
present. The overall effect is a change in the equation
of motion for the particle. The nature of this modified equation
has been subject to much controversy since the initial work of
Abraham and Lorentz. The standard equation for the change in the
energy and momentum of the particle in classical electrodynamics
is the relativistic generalization by Dirac of the work of
Abraham and Lorentz and called
the Lorentz-Dirac (or Abraham-Lorentz-Dirac) equation
~\cite{Abraham,Lorentz,Dirac}.  (See, e.g., Ref.~\cite{Poisson}
for a modern review.) Thus, if a charge $e$ with mass $m$ is
accelerated by an external 4-force $F^\mu_{\rm ext}$, then its
coordinates $x^\mu(\tau)$ at the proper time $\tau$ obey the
following equation:
\begin{equation}
m\frac{d^2x^\mu}{d\tau^2} = F^\mu_{\rm ext} + F^\mu_{\rm LD}\,,
\label{Lor}
\end{equation}
where the Lorentz-Dirac 4-force $F^\mu_{\rm LD}$ is given by
\begin{equation}\label{LD4force}
F^\mu_{\rm LD} \equiv \frac{2\alpha_c}{3}\left[
\frac{d^3x^\mu}{d\tau^3} + \frac{dx^\mu}{d\tau}\left(\frac{d^2
x^\nu}{d\tau^2} \frac{d^2x_\nu}{d\tau^2}\right)\right]\,.
\end{equation}
We have let $c=1$ and defined $\alpha_c\equiv e^2/4\pi$. Our
metric is $g_{\mu\nu} = {\rm diag}\,(+1,-1,-1,-1)$.  This equation
of motion is fundamentally different in form from those which one
usually encounters in mechanics
as it is third-order. This implies that a third
initial condition is needed in addition to the position
and velocity.  Herein lies the cause of much of the debate as to
the physical correctness of this theory. It should be noted that
there are many derivations of the Lorentz-Dirac force
using a number of different methods besides the original work of
Dirac~\cite{Dirac} (see, e.g., Ref.~\cite{Teit}).
However, many unphysical and problematic
effects occur in the
implementation. One example is the existence of run-away solutions
for which a particle will continue to accelerate under its own
radiation reaction. This is certainly an unphysical effect and
thus a problem for the theory. Alternative forms of implementing the
Lorentz-Dirac force, including the integro-differential form,
solve the problem of run-aways but create another problem in
the form of pre-acceleration. Thus, in this implementation of the
Lorentz-Dirac force the particle
accelerates before the external
force causing the acceleration is applied; this is
an acausal effect which should not be present in classical mechanics.
This effect is represented in Fig.
\ref{preacceleration}, which shows
the acceleration of the particle, $a$, against
time, $t$, where a constant
force is applied after time $t_0$, i.e. in the
shaded region.

Additional conditions and
procedures are therefore necessary in order to treat the
Lorentz-Dirac theory as a a normal causal classical theory, and
there are such procedures (see, for example, the review
\cite{Poisson} or the discussions in \cite{Jackson} and
\cite{Rohrlich}). One such procedure is the
{\it reduction of order}~\cite{Landau}, in which the Lorentz-Dirac force
is treated as a perturbation order by order.
Despite this, the debate over the controversial aspects of
Eq.~(\ref{LD4force}) has continued, its impact
muted for a couple of reasons. Firstly, the
effect due to the third-derivative term, often called the Schott term,
is tiny for most physical situations; in fact the
time scale for the pre-acceleration is too small for
any known classical interactions for it to have a measurable effect.
Secondly, the advent of quantum
electrodynamics (QED), which is the more fundamental theory
of electrodynamics, has rendered the problems concerning the
classical Lorentz-Dirac force less urgent.  The second point, however,
naturally leads to the following question: ``If the perturbative
treatment of the Lorentz-Dirac force is causal and satisfactory,
can it be derived in QED in the $\hbar \to 0$ limit in perturbation
theory?"  This question
is the main motivation of this paper.

In fact it was found in Refs.~\cite{Higuchi1,Higuchi2}
(after an initial claim to the contrary)
that the change of the position, called the position shift,
due to the Lorentz-Dirac force
of a charged particle linearly accelerated by
a static external potential is reproduced by the $\hbar \to 0$
limit of one-photon emission process in QED in the
non-relativistic approximation. (See, e.g.,
Refs.~\cite{MS,beilok,tsyto}
for other approaches to arrive at the Lorentz-Dirac theory from
QED.) Here we present in detail the generalization of this work to a
fully relativistic particle, which was outlined in
\cite{HM}. At the end of the paper we shall also show similar
results with a time-dependent but space-independent potential. The
paper is organized as follows. In Secs.~\ref{Cmodel} and
\ref{Qmodel} we outline the classical and quantum models that are
used in the paper. In Sec.~\ref{PEV} we calculate expressions for
the position expectation value for both the hypothetical
non-radiating particle (Subsection~\ref{nonradPEV}) and the
radiating particle (Subsection~\ref{radPEV}) before identifying
the expression for the position shift in terms of the emission
amplitude. This amplitude is calculated in Sec.~\ref{ampcalc}. The
result is used in Sec.~\ref{larmorsec} to show that the energy
emitted in the $\hbar
\to 0$ limit in QED is given by the classical Larmor formula and to compute the quantum position shift in
Sec.~\ref{qposshiftcalc}. The comparison between the classical and
quantum position shifts is presented in detail in
Sec.~\ref{comparison}  and the time-dependent case is presented in
Sec.~\ref{timeextension} before we conclude the paper in
Sec.~\ref{conclusion}.

\section{Classical Model}\label{Cmodel}

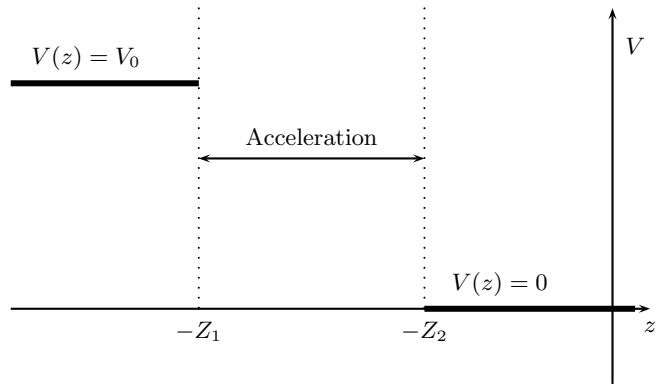
\begin{figure}
\begin{center}
\begin{pspicture}(0.5,-1)(9,4)
 \psline{->}(0.5,0)(9,0) \psline{->}(8.5,-1)(8.5,4)
 \psline[linewidth=0.8mm](0.5,3)(3,3)
 \psline[linewidth=0.8mm](6,0)(8.8,0)
 \uput[r](8.5,3.5){$V$} \uput[d](9,0){$z$}
 \psline[linestyle=dotted](3,0)(3,4)
 \psline[linestyle=dotted](6,0)(6,4)
 \uput[u](1.5,3){$V(z)=V_0$} \uput[u](7,0){$V(z)=0$}
 \uput[d](3,0){$-Z_1$} \uput[d](6,0){$-Z_2$}
 \psline{<->}(3,2)(6,2)
 \uput[u](4.5,2){Acceleration}
\end{pspicture}
\end{center}
\caption{The potential $V(z)$ and period of acceleration}
\label{potentialmodel}
\end{figure}

We start by describing the model to be investigated in this paper.
Consider a charged particle with charge $e$ and mass $m$ moving in
one space dimension under a potential. Let this motion be in the
positive $z$-direction and let the potential be a function of the
spatial coordinate $z$, viz $V=V(z)$. We wish to analyze the change in
the position of the particle due to radiation reaction. Thus we
consider a model in which there has been a period of
acceleration, i.e. non-constant potential, at some time in the
particle's history. We assume that $V(z)=V_0=
\text{const.}$ for $z<-Z_1$ and $V(z)=0$ for $-Z_2<z$ for some
$Z_1$ and $Z_2$, both positive constants with $Z_1>Z_2$.
Thus, there is
non-zero acceleration only in the interval $(-Z_1,-Z_2)$ as
represented in Fig. \ref{potentialmodel} for the case with $V_0 > 0$.

The external 4-force in Eq.~(\ref{Lor}) representing the electric force
corresponding to this potential is given as
\begin{eqnarray}
 F^t_{\rm ext} &=&-V'(z)\,dz/d\tau \,,  \nonumber \\
 F^z_{\rm ext} &=& - V'(z)\,dt/d\tau \,, \nonumber \\
 F^x_{\rm ext} &=& F^y_{\rm ext} = 0 \,.
\end{eqnarray}
The Lorentz-Dirac 4-force can be given in the following form:
\begin{eqnarray}
 F^t_{\rm LD} &=& F_{\rm LD}\,dz/d\tau \nonumber \,, \\
 F^z_{\rm LD} &=& F_{\rm LD}\,dt/d\tau \nonumber \,, \\
 F^x_{\rm LD} &=& F^y_{\rm LD} = 0
\end{eqnarray}
with
\begin{equation}
F_{\rm LD}  \equiv \frac{2\alpha_c}{3}\gamma \frac{d\ }{dt}
(\gamma^3\ddot{z})\,.  \label{LDLD}
\end{equation}
A dot indicates the derivative with respect to $t$. We have
defined $\gamma \equiv (1-\dot{z}^2)^{-1/2}$ as usual.

Suppose that this particle would be at $z=0$ at time $t=0$ {\em if
the Lorentz-Dirac force was absent}.  The true position at $t=0$,
which we denote $\delta z$ and call the position shift, can
readily be found to lowest non-trivial order in $F_{\rm LD}$ by
treating the Lorentz-Dirac force as perturbation. The calculation
can be facilitated by using the fact that the total energy,
$m\,dt/d\tau + V(z)$, changes by the amount of work done by the
Lorentz-Dirac force. Thus we find
\begin{eqnarray}\label{0}
 \int_{-\infty}^{t} F_{\rm LD}(t')\dot{z}(t') dt'
&=&\frac{d}{d\dot{z}}\frac{m}{\sqrt{1-\dot{z}^2}} \delta \dot{z}
 + V'(z)\delta z \nonumber \\
&=&  m\gamma^3\dot{z}^2\frac{d}{dt}\left(\frac{\delta
  z}{\dot{z}}\right)\,,
\end{eqnarray}
where we have used
\begin{equation}\label{vdash}
\frac{d\ }{dt}\left( m\gamma \dot{z}\right)
= m \gamma^3\ddot{z} = - V'(z)
\end{equation}
to zeroth order, i.e. in the absence of the Lorentz-Dirac force.
Rearranging
and integrating, and then interchanging the order of integration, we
obtain the position shift as
\begin{equation}
\delta z_{\rm LD} = -\frac{v_0}{m}\int^0_{-\infty}
\left(\int^{t}_{0}\frac{1}{\gamma^3(t')[\dot{z}(t')]^2}dt'
\right)F_{\rm LD}\frac{dz}{dt}dt\,,\label{deltaz}
\end{equation}
where $v_0 = \dot{z}|_{t=0}$ is the final velocity in the absence
of radiation reaction. We
shall see
that this result is reproduced in QED in the $\hbar \to 0$
limit.

It is useful for later purposes to find the change in the position
shift (\ref{deltaz}) caused by letting the particle be at
$z=z_0\neq 0$ at $t=0$ with the same final
velocity $v_0$ in the absence of the Lorentz-Dirac force.
(We assume that $z_0 > - Z_2$.)
The time this particle spends after the acceleration
and before getting to $t=0$ (when the position shift is ``measured")
is lengthened by $z_0/v_0$.  Hence this position shift is obtained by
using the same trajectory but defining it at
$t_0\equiv z_0/v_0$.  Hence, writing this position shift as
\begin{equation}
\delta z_{\rm class} = \delta z_{\rm LD} + \delta z_{\rm extra}\,,
\label{classicalone}
\end{equation}
we find
\begin{equation}\label{zextra}
\delta z_{\rm extra} =
-\frac{v_0}{m}\int_{-\infty}^{t_0}
\left( \int_{t_0}^0 \frac{1}{\gamma^3(t')\left[\dot{z}(t')\right]^2}
dt'\right)F_{\rm LD}\frac{dz}{dt}dt\,.
\end{equation}
Noting that $\dot{z}(t)=v_0$ for values of $t$
between $0$ and $t_0$ we obtain
\begin{equation}
\delta z_{\rm extra}
=  -\frac{z_0}{m\gamma_0^3 v_0^2}E_{\rm em} \label{extraaa}
\end{equation}
with $\gamma_0 \equiv (1-v_0^2)^{-1/2}$, where $E_{\rm em}$ is the
energy emitted as radiation given by
\begin{eqnarray}
E_{\rm em} & = & - \int_{-\infty}^0 F_{\rm LD}\frac{dz}{dt}dt
\nonumber \\
& = & \frac{2\alpha_c}{3}\int_{-\infty}^0 (\gamma^3 \ddot{z})^2\,dt\,,
\label{Larmor}
\end{eqnarray}
which is the relativistic Larmor formula for one-dimensional motion.
We shall see that the
extra contribution (\ref{extraaa}) to the position shift is also
reproduced by QED in the $\hbar \to 0$ limit.

\section{The QED Model}\label{Qmodel}

We now turn our attention to the quantum field theoretic model.
The corresponding Lagrangian density is
\begin{eqnarray}
{\cal L} & = &
[(D_\mu + ieA_\mu)\varphi]^\dagger [(D^\mu + ie A^\mu)\varphi]
- (m/\hbar)^2\varphi^\dagger \varphi \nonumber \\
&& - \frac{1}{4}F_{\mu\nu}F^{\mu\nu}
- \frac{1}{2}(\partial_\mu A^\mu)^2\,,\label{firstLag}
\end{eqnarray}
where $F_{\mu\nu} \equiv \partial_\mu A_\nu - \partial_\nu A_\mu$
and $D_\mu \equiv \partial_\mu + i V_\mu/\hbar$. The
field $\varphi$ describes a charged scalar particle with mass $m$
and charge $e$, the field $A_\mu$ is the electromagnetic field,
and the function $V_\mu$ is the external potential which
accelerates the charged scalar particle and is given as
$V_\mu=V(z)\delta_{\mu 0}$ for the static potential. The
non-interacting quantum
electromagnetic field $A_\mu(x)$ is expanded as
\begin{equation}
A_\mu(x) = \int\frac{d^3{\bf k}}{2k(2\pi)^3} \left[ a_\mu({\bf k})
e^{-ik\cdot x} + a_\mu^\dagger({\bf k})e^{ik\cdot x} \right]\,,
\end{equation}
where $k = \|{\bf k}\|$ and the annihilation and creation operators,
$a_\mu({\bf k})$ and $a^\dagger_\mu({\bf k})$, respectively,
for the photons with momenta $\hbar {\bf k}$
in the Feynman gauge satisfy
\begin{equation}\label{commu1}
\left[ a_\mu({\bf k}),a^\dagger_\nu({\bf k}')\right] =
-g_{\mu\nu}(2\pi)^32\hbar k\delta({\bf k}-{\bf k}')\,.
\end{equation}

The Fourier expansion of the non-interacting quantum scalar field
is not as straightforward because its
field equation involves the potential $V(z)$ as follows:
\begin{equation}
\left\{ \hbar^2[i\partial_t - V(z)]^2 + \hbar^2\nabla^2 -
m^2\right\}\varphi = 0\,.
\end{equation}
Positive-frequency mode functions can be found in the form
\begin{equation}\label{LargePhi}
\Phi_{\bf p}(t,{\bf x}) = \phi_p(z)\exp\left[\frac{i}{\hbar}
({\bf p}_\perp\cdot {\bf x}_\perp - p_0 t)\right]\,,\ \
p_0 > 0\,,
\end{equation}
where ${\bf x}_\perp \equiv (x,y)$ and
${\bf p}_\perp \equiv (p^x,p^y)$.  We {\em define} the $z$-component
of the momentum $p \equiv p^z$ by the relation
$p_0^2 = p^2 + {\bf p}_\perp^2 + m^2$.  We let $p$ be positive
(negative) if the mode function corresponds to a wave coming in from
$z=-\infty$ ($z=+\infty$).
The function $\phi_p(z)$ with ${\bf p}_\perp = 0$ satisfies
\begin{equation}
-\hbar^2 \frac{d^2\ }{dz^2}\phi_p(z) = [\kappa_p(z)]^2\,\phi_p(z)\,,
\end{equation}
where
\begin{equation}\label{kappap}
\kappa_p(z) \equiv  \sqrt{[p_0 - V(z)]^2 - m^2}\,.
\end{equation}
If ${\bf p}_\perp \neq 0$, then the mass term $m^2$ has to be
replaced by $m^2 + {\bf p}_\perp^2$.  However, we need only the cases
with ${\bf p}_\perp=0$ and $p > 0$ in our calculations.
The WKB approximation for the function $\phi_p(z)$ with $p>0$,
which will be useful later, reads
\begin{equation}\label{phipp}
\phi_p(z) = \sqrt{\frac{p}{\kappa_p(z)}}\,
\exp\left[\int_0^z \kappa_p(\zeta)\,d\zeta\right]\,,
\end{equation}
where we have required $\phi_p(0) = 1$.

The non-interacting quantum charged scalar field is expanded in
terms of these mode functions as
\begin{equation}\label{expand}
\varphi(x) = \hbar \int \frac{d^3{\bf p}}{2p_0(2\pi\hbar)^3}
\left[ A({\bf p})\Phi_{\bf p}(x) + B^\dagger({\bf p}) \Phi^*_{\bf
p}(x)\right]\,.
\end{equation}
(This expansion needs to be modified if there are modes which
are totally reflected by the potential --- this is the case if
$V_0 \neq 0$ --- or if there are
modes which fail to reach $|z|=\infty$.  However, such modifications
will not be relevant since the affected modes will not enter
our calculations.)
The operators $A({\bf p})$ and $B({\bf p})$ satisfy
\begin{equation}\label{commu2}
\left[ A({\bf p}),A^\dagger({\bf p}')\right] =
\left[ B({\bf p}),B^\dagger({\bf p}')\right] =
(2\pi\hbar)^32p_0\delta({\bf p}-{\bf p}')\,,
\end{equation}
if one normalizes the mode functions appropriately,
with all other commutators vanishing.  The WKB mode functions
$\Phi_{\bf p}(t,{\bf x})$ with $\phi_p(z)$ given by Eq.~(\ref{phipp})
are correctly normalized since it satisfies the usual normalization
condition
$|\Phi_{\bf p}(t,{\bf x})| = 1$ in the region $(-Z_2,+\infty)$ where
$V(z)=0$.

Now, in the interaction picture,
if there is one scalar particle with momentum ${\bf p}$ in the
initial state, then the final state to lowest non-trivial order in $e$
is a superposition of a state
proportional to the initial
state and states with one scalar particle and one photon.  Thus, the
state $A^\dagger({\bf p})|0\rangle$ evolves as follows:
\begin{eqnarray}
A^\dagger({\bf p})|0\rangle & \to & [1 + i{\cal F}({\bf
p})]A^\dagger({\bf p})|0\rangle\nonumber \\ &&  + \frac{i}{\hbar}
\int \frac{d^3{\bf k}}{2k(2\pi)^3} {\cal A}^\mu({\bf p},{\bf
k}) a_\mu^\dagger({\bf k})A^\dagger({\bf P})|0\rangle\,. \nonumber
\\ \label{trans}
\end{eqnarray}
The ${\cal F}({\bf p})$ is the forward-scattering amplitude
of order $e^2$ coming
from the one-loop diagram, which we do not evaluate
explicitly. The emission amplitude ${\cal A}_\mu({\bf p},{\bf k})$
will play a central role in our calculations.
Note that the momentum $\hbar {\bf k}$ of the photon
is of order $\hbar$ because the wave number ${\bf k}$ rather than
the momentum has the classical limit.  The momentum ${\bf P}$ of
the charged particle in the final state is determined using
conservation of the transverse momentum and energy:
\begin{eqnarray}
{\bf p}_\perp & = & {\bf P}_\perp + \hbar {\bf k}_\perp\,,\\
p_0 & = & P_0 +\hbar k\,.
\end{eqnarray}
The $z$-component $P\equiv P^z$ is determined by the on-shell condition
\begin{equation}
P_0^2 = P^2 + {\bf P}_\perp^2 + m^2\,.
\end{equation}

We shall consider the initial state $|i\rangle$ given by
\begin{equation}\label{istate}
| i\rangle = \int \frac{d^3{\bf p}}{\sqrt{2p_0}(2\pi\hbar)^3}
f({\bf p})A^\dagger({\bf p})|0\rangle\,,
\end{equation}
where the function $f({\bf p})$ is sharply peaked about a momentum
in the positive $z$-direction with width of order $\hbar$.  We also
require that the $p$-derivative of $f({\bf p})$ has width of order
$\hbar$.  This wave packet corresponds to the classical particle
considered in the previous section.
We make a further assumption
that the WKB approximation (\ref{phipp}) is valid for the
momenta ${\bf p}$ where the function $f({\bf p})$ is not negligibly
small. The WKB approximation is known to be applicable
if the wavelength stays
approximately constant over many periods
(see, e.g. Ref.~\cite{Shankar}).  Since the ``time-dependent
wavelength" is $2\pi\hbar[\kappa_p(z)]^{-1}$,
where $\kappa_p(z)$ is defined by
Eq.~(\ref{kappap}), this condition is
\begin{equation}
\hbar\left|\frac{d\ }{dz}\left[\frac{1}{\kappa_p(z)}\right]\right|
\ll 1\,,
\end{equation}
which is automatically satisfied as long as
$\kappa_p(z)$ and $\kappa'_p(z)$ can be regarded
as quantities of order $\hbar^0$.

The normalization
condition $\langle i\,|\,i\rangle = 1$ implies
\begin{equation}\label{inorm}
\int\frac{d^3{\bf p}}{(2\pi\hbar)^3}|f({\bf p})|^2 = 1
\end{equation}
because of the commutation relations (\ref{commu2}).
This equation shows that the function $f({\bf p})$ can heuristically
be regarded as the
one-particle wave function in the momentum representation.  In
Ref.~\cite{HM} it was assumed that $f({\bf p})$ was real for simplicity.
This assumption will be dropped in this paper.

\section{Position expectation value}\label{PEV}

In this section we derive a formula for the position shift in terms of
the emission amplitude ${\cal A}^\mu({\bf p},{\bf k})$.  We closely
follow Ref.~\cite{Higuchi2}, correcting a few errors.
In particular, the fourth term in Eq.~(\ref{zsum}) was missing in
Ref.~\cite{Higuchi2}.

\subsection{Non-radiating particle}\label{nonradPEV}

Define the charge density by
\begin{equation}
\rho(x) \equiv\frac{i}{\hbar}:\varphi^\dagger\partial_t \varphi -
\partial_t \varphi^\dagger\cdot \varphi:\,, \label{charge}
\end{equation}
where $:\ldots:$ denotes normal ordering.
The operator $\rho(x)$ is the $t$-component of the conserved current
\begin{equation}
  J^\mu(x) \equiv \frac{i}{\hbar} : \varphi^\dagger\partial^\mu\varphi -
\partial^\mu\varphi^\dagger\cdot \varphi:\,.
\end{equation}
We let the potential satisfy $|V_0| < 2m$, thus precluding
the possibility of scalar-particle pair creation.
Then, the charge density $\rho(x)$
coincides with the probability density for the particle if
there is only one charged particle in the state.
Hence the expectation value of $z$ is
\begin{equation}
\langle z\rangle = \int d^3{\bf x}\, z\,\langle \rho(t,{\bf x})
\rangle \,. \label{zedd}
\end{equation}
We shall first need to consider the expectation value of $z$
for the hypothetical scalar particle that interact with the external
potential but not with the quantum electromagnetic field. This
will be the benchmark against which we can define the position
shift due to the radiation. Hence,
we must evaluate the expectation value
of the probability density in the absence of radiation and
use Eq.~(\ref{zedd}) to find $\langle z\rangle$. The
initial and final states in this case are the same.  Thus,
using the commutation relations
(\ref{commu2}) for the expectation value of the charge
density (\ref{charge})
in the state $|i\rangle$ given by (\ref{istate}), we
find
\begin{eqnarray}
&& \langle \rho(t,\bf{x})\rangle\nonumber \\
&& = i\hbar \int\frac{d^3{\bf
p}'}{\sqrt{2p_0'}(2\pi\hbar)^3}
 \int\frac{d^3{\bf p}}{\sqrt{2p_0}(2\pi\hbar)^3}
f^*({\bf p}') f({\bf p})\nonumber \\
&& \times \left[\Phi_{{\bf
p}'}^*(t,{\bf x})\partial_t \Phi_{{\bf p}}(t,{\bf x}) -
\partial_t\Phi_{{\bf p}'}^*(t,{\bf x})\cdot \Phi_{{\bf p}}(t,{\bf
x})\right]\,. \nonumber \\
\end{eqnarray}
The $t$-dependence of the mode functions is given by
$\Phi_{\bf p}(t,{\bf x}) \propto
e^{-i p_0 t /\hbar}$.  Hence, we find
\begin{eqnarray}\label{rho}
\langle \rho(t,{\bf x})\rangle & = & \frac{1}{2}
\int\frac{d^3{\bf p}'}{(2\pi\hbar)^3}
 \int\frac{d^3{\bf p}}{(2\pi\hbar)^3}f^*({\bf p}') f({\bf p})
\nonumber \\
&& \times
 \left( \sqrt{\frac{p_0}{p_0'}} +
\sqrt{\frac{p_0'}{p_0}}\right)\Phi_{{\bf p}'}^*(t,{\bf x})
\Phi_{{\bf p}}(t,{\bf x})\,.  \nonumber \\
\end{eqnarray}
We arrange the wave packet so that at time $t=0$
it is located far into
the region where the potential vanishes. Then we may approximate
the mode functions as follows:
\begin{equation}\label{appp}
  \Phi_{\bf p}(0,{\bf x}) \approx
e^{-i(p z + \bf{p}_{\perp}\cdot \bf{x}_{\perp})/\hbar}\,.
\end{equation}
After substituting Eq.~(\ref{rho}) in the position
expectation formula (\ref{zedd}) we use the
approximation (\ref{appp}) to write
\begin{equation}\label{zfromphi}
z\Phi_{\bf p}(0,{\bf x}) \approx -i \hbar \frac{\partial}{\partial p}
\Phi_{\bf p}(0,{\bf x})\,.
\end{equation}
We can now  use integration by parts for the variable $p$ in
Eq.~(\ref{zedd}) to obtain the expectation value of $z$ for
the hypothetical non-radiating particle, which we denote by $z_0$, as
\begin{eqnarray}
z_0 &=& \frac{i\hbar}{2}\int\frac{d^3{\bf
p}'}{(2\pi\hbar)^3}
 \int\frac{d^3{\bf p}}{(2\pi\hbar)^3} \nonumber \\
  & \times & \frac{\partial}{\partial p}
\left[ \left( \sqrt{\frac{p_0}{p_0'}} +
\sqrt{\frac{p_0'}{p_0}}\right)f^*({\bf p}') f({\bf p}) \right]
\nonumber
\\ & \times & \int d^3{\bf x}\,
\Phi_{{\bf p}'}^*(0,{\bf x}) \Phi_{{\bf p}}(0,{\bf x})\,.
\end{eqnarray}
Substituting the approximation (\ref{appp}) for $\Phi_{\bf p}$
gives a delta function:
\begin{eqnarray}
z_0 &=& \frac{i\hbar}{2}\int\frac{d^3{\bf p}'}
{(2\pi\hbar)^3}
 \int\frac{d^3{\bf p}}{(2\pi\hbar)^3}f^*({\bf p}') \nonumber \\
  & \times &  \frac{\partial}{\partial p}
\left[ \left( \sqrt{\frac{p_0}{p_0'}} +
\sqrt{\frac{p_0'}{p_0}}\right) f({\bf p}) \right]
\nonumber \\ & \times & (2\pi\hbar)^3 \delta(\bf{p}-\bf{p'})\,.
\end{eqnarray}
Performing the ${\bf p}'$-integration after explicitly working out
the $p$-derivative, we find
\begin{eqnarray}
z_0 &=& i\hbar\int\frac{d^3{\bf p}}
{(2\pi\hbar)^3}f^*({\bf p}) \frac{\partial}{\partial p}
 f({\bf p})
 \end{eqnarray}
or, by integration by parts,
\begin{equation}\label{nonradshift}
z_0 = \frac{i\hbar}{2} \int\frac{d^3{\bf
p}}{(2\pi\hbar)^3} f^*({\bf p}) \stackrel{\leftrightarrow}{\partial}_p
f({\bf p})\,,
\end{equation}
where $\stackrel{\leftrightarrow}{\partial}_p =
\stackrel{\rightarrow}{\partial}_p - \stackrel{\leftarrow}
{\partial}_p$.  This result can be interpreted as the expectation
value of the position operator $i\hbar\partial_p$ in the
${\bf p}$-representation of the one-particle wave function.
An example of a function $f({\bf p})$ satisfying
Eq.~(\ref{nonradshift}) is
$f({\bf p}) = f_R({\bf p})e^{-iz_0p/\hbar}$ with the function
$f_R({\bf p})$ being real though it is not
necessary to make this assumption.

\subsection{Radiating particle}\label{radPEV}

Now we consider the case of the radiating particle. The final state
resulting from the initial state $|i\rangle$ can be found from
Eq.~(\ref{trans}) as
\begin{eqnarray}\label{final1}
| f\rangle & = & \int\frac{d^3{\bf p}}{\sqrt{2p_0}(2\pi\hbar)^3}
F({\bf p}) A^\dagger({\bf p})|0\rangle \nonumber
\\
&& + \frac{i}{\hbar}\int \frac{d^3{\bf k}}{2k(2\pi)^3}
\int\frac{d^3{\bf p}}{\sqrt{2p_0}(2\pi\hbar)^3}\nonumber \\ && \ \
\ \ \ \ \ \ \ \times G^\mu({\bf p},{\bf k}) a_\mu^\dagger({\bf
k})A^\dagger({\bf P})|0\rangle\,,
\end{eqnarray}
where ${\bf P}_\perp = {\bf p}_\perp - \hbar {\bf k}_\perp$ and
$P_0 = p_0 - \hbar k$ as before.
We have defined
\begin{eqnarray}\label{Fdefn}
F({\bf p}) \equiv \left[ 1 + i{\cal F}({\bf p})
\right]f({\bf p})\,,  \\
G^\mu({\bf p},{\bf k}) \equiv {\cal A}^\mu({\bf p},{\bf k})f({\bf
p})\,. \label{Gdefn}
\end{eqnarray}
One can heuristically
regard the function $F({\bf p})$ as the one-particle wave
function in the zero-photon sector in the ${\bf p}$-representation
and the function $G^\mu({\bf p},{\bf k})$ as that in the
one-photon sector with a photon with momentum $\hbar {\bf k}$ in
the ${\bf P}$-representation. By introducing the definition
\begin{equation}
C^{\dagger\mu}({\bf k}) = \frac{i}{\hbar}\int\frac{d^3{\bf
p}}{\sqrt{2p_0}(2\pi\hbar)^3} {\cal A}^\mu({\bf p},{\bf k})f({\bf
p})A^\dagger({\bf P})\,, \label{diff}
\end{equation}
the final state can be written
\begin{eqnarray}\label{final2}
| f\rangle &=&  \int\frac{d^3{\bf p}}{\sqrt{2p_0}(2\pi\hbar)^3}
F({\bf p}) A^\dagger({\bf p})|0\rangle \nonumber
\\ && + \int \frac{d^3{\bf k}}{2k(2\pi)^3} C^{\dagger\mu}({\bf
k})a_\mu^{\dagger}({\bf k})|0\rangle\,.
\end{eqnarray}

We first derive the relation between the imaginary part of
${\cal F}({\bf p})$ and the scattering probability which results
from unitarity.
Recalling the definition (\ref{Fdefn}) and using the
unitarity of time evolution, we find
\begin{equation}
1 = \langle f\,|\,f\rangle = \int\frac{d^3{\bf p}}{(2\pi)^3} (1 -
2\,{\rm Im}\,{\cal F}({\bf p}))|f({\bf p})|^2 +{\cal P}_{\rm
em}\,,  \label{unit}
\end{equation}
to first order in $e^2$, where the emission probability ${\cal
P}_{\rm em}$ is given by
\begin{eqnarray}
{\cal P}_{\rm em}
& = & - \hbar \int \frac{d^3{\bf k}}{2k(2\pi)^3} \langle 0
|C_{\mu}({\bf k}') C^{\dagger\mu}({\bf k})|0\rangle\nonumber \\ &
= & - \frac{1}{\hbar}\int \frac{d^3{\bf k}}{2k(2\pi)^3}
\int \frac{d^3{\bf p}}{\sqrt{2p_0}(2\pi\hbar)^3} f^*({\bf p}){\cal A}_\mu^*
({\bf p},{\bf k})\nonumber \\
&& \times \int\frac{d^3{\bf p}'}{\sqrt{2p_0'}(2\pi\hbar)^3}
f({\bf p}'){\cal A}^\mu({\bf p}',{\bf k})\nonumber \\
&& \times \langle 0|A({\bf P})A^\dagger({\bf P}')|0\rangle\,.
\label{abc}
\end{eqnarray}
To perform the ${\bf p}'$-integration
recall first that
\begin{equation}
 \langle 0|A({\bf P})A^\dagger({\bf P}')|0\rangle = (2\pi \hbar)^3
2P_0\delta({\bf P} - {\bf P}')\,.  \label{delPP}
\end{equation}
We need to integrate this delta function with respect to ${\bf p}'$
by relating it to $\delta({\bf p}-{\bf p}')$.
We note that
\begin{equation}
 \delta({\bf P}_\perp - {\bf
P}'_\perp) = \delta({\bf p}_\perp - {\bf p}'_\perp)
\end{equation}
because ${\bf P}_\perp = {\bf p}_\perp - \hbar{\bf k}_\perp$, and
\begin{equation}
 \delta(P - P') = \frac{dp}{dP}\delta(p-p')\,.
\end{equation}
By using these formulae in Eq.~(\ref{delPP}) we find
\begin{equation}
\langle 0|A({\bf P})A^\dagger({\bf P}')|0\rangle
= \frac{P_0}{p_0}\frac{dp}{dP}(2\pi\hbar)^32p_0\delta
({\bf p}-{\bf p}')\,.
\end{equation}
By substituting this formula in Eq.~(\ref{abc}) we obtain
\begin{eqnarray}\label{emprobtrue}
{\cal P}_{\rm em} & = &
- \frac{1}{\hbar}\int \frac{d^3{\bf p}}{(2\pi\hbar)^3}
|f({\bf p})|^2 \nonumber \\
&& \times
\int \frac{d^3{\bf k}}{2k(2\pi)^3}
{\cal A}_\mu^*({\bf p},{\bf k}){\cal A}^\mu({\bf p}',{\bf k})
\frac{P_0}{p_0}\frac{dp}{dP}\,.  \nonumber \\
\label{thiseq}
\end{eqnarray}
Eq.\ (\ref{unit})
must hold for any function $f({\bf p})$, so we must have
\begin{equation}
2\,{\rm Im}{\cal F}({\bf p}) = -\frac{1}{\hbar}
\int\frac{d^3{\bf k}}{2k(2\pi)^3}
{\cal A}_\mu^*({\bf p},{\bf k}){\cal A}^\mu({\bf p},{\bf k})
\frac{P_0}{p_0}\frac{dp}{dP}\,. \label{unitary}
\end{equation}

The contribution of the one-photon-emission term to $\langle
\rho(t,{\bf x})\rangle$ is
\begin{equation}
\langle \rho(t,{\bf x})\rangle_{\rm em}
=  -\hbar\int\frac{d^3{\bf
k}}{2k(2\pi)^3} \langle 0|C^\mu({\bf k})\rho(t,{\bf
x})C_\mu^\dagger({\bf k})|0\rangle\,.
\label{one-photon}
\end{equation}
The integrand here is obtained from  $\langle
\rho(t,{\bf x})\rangle$ for the non-radiating particle if we
replace the state $|i\rangle$ in Eq.~(\ref{istate}) by
\begin{equation}
\hbar^{1/2}C_\mu^\dagger({\bf k})|0\rangle
=  i \int\frac{d^3{\bf P}}{\sqrt{2P_0}(2\pi\hbar)^3}\,g_\mu({\bf P})
A^\dagger({\bf P})|0\rangle\,,
\end{equation}
where
\begin{equation}
g_\mu ({\bf P}) \equiv \hbar^{-1/2}
{\cal A}_\mu({\bf p},{\bf k})f({\bf p})
\sqrt{\frac{P_0}{p_0}}\,\frac{dp}{dP}\,.
\end{equation}
Hence we conclude from Eq.~(\ref{nonradshift}) that
\begin{eqnarray}
&& \hbar \int d^3{\bf x}\, z\,\langle 0| C^\mu({\bf k})\rho(t,{\bf
x})C^\dagger_\mu ({\bf k})|0\rangle\nonumber \\
&& = \frac{i\hbar}{2}\int\frac{d^3{\bf P}}{(2\pi)^3}\,
g^{\mu *}({\bf P})\stackrel{\leftrightarrow}{\partial_P} g_\mu({\bf P})\,.
\end{eqnarray}
This implies
that the one-photon-emission contribution to $\langle z\rangle$ is
\begin{eqnarray}
\langle z \rangle_{\rm em} &=& -\frac{i}{2}
\int \frac{d^3{\bf
k}}{2k(2\pi)^3}
\int\frac{d^3{\bf P}}{(2\pi \hbar)^3}
\nonumber \\ && \times \left[f^*({\bf p}) {\cal
A}^{\mu *}({\bf p},{\bf k})\right]
\stackrel{\leftrightarrow}{\partial}_P\left[ f({\bf p}){\cal
A}_\mu({\bf p},{\bf k})\right]\nonumber \\
&& \times \frac{P_0}{p_0}\left(\frac{dp}{dP}\right)^2\,.
\end{eqnarray}
We convert the integration
variable from ${\bf P}$ to ${\bf p}$ by using
$d^3{\bf P} = (\partial P/\partial p)d^3{\bf p}$ and the
$P$-derivative to a $p$-derivative by using
$\partial_P = (\partial p/\partial P)\partial_p$ to find
\begin{eqnarray}
\langle z  \rangle_{\rm em}  & = &
-\frac{i}{2} \int \frac{d^3{\bf p}}{(2\pi\hbar)^3} f^*({\bf
p})\stackrel{\leftrightarrow}{\partial}_p f({\bf p})
 \nonumber \\
&& \times \int\frac{d^3{\bf k}}{2k(2\pi)^3} {\cal A}^{\mu
*}({\bf p},{\bf k}){\cal A}_\mu({\bf p},{\bf k})
 \frac{P_0}{p_0}\left(\frac{dp}{dP}\right)^2
\nonumber \\ && -
\frac{i}{2} \int \frac{d^3{\bf p}}{(2\pi\hbar)^3} | f({\bf
p})|^2
\nonumber \\ && \times \int \frac{d^3{\bf k}}{2k(2\pi)^3}
{\cal A}^{\mu *}({\bf p},{\bf k})
\stackrel{\leftrightarrow}{\partial}_p {\cal A}_\mu({\bf p},{\bf
k})\,.
\end{eqnarray}
We have dropped the factor $(P_0/p_0)(dp/dP)^2$ from the second
term for the following reason.  It will turn out later that
the emission amplitude ${\cal A}^{\mu}$ and its $p$-derivative
are both of order $\hbar^0$.
Thus the second term in the above expression for $\langle z
\rangle_{\rm em}$ is of order $\hbar^0$.
Since $P_0=p_0$ and $dp/dP=1$
to order $\hbar^0$, we can drop the factors of $P_0/p_0$ and
$dp/dP$ from the second term.
However, one needs to keep these factors
in the first term, which is of order $\hbar^{-1}$.

{}From the expressions for the final state (\ref{final2}) it is
clear that the contribution to $\langle z \rangle$ from the
process without photon emission is obtained from
Eq.~(\ref{nonradshift}) by replacing $f(\bf{p})$ with $F({\bf p})
\equiv [1 + i{\cal F}({\bf p})]f({\bf p})$. Thus
\begin{eqnarray}
\langle z\rangle_{\rm for}
&= &  \frac{i\hbar}{2}
\int\frac{d^3{\bf p}}{(2\pi\hbar)^3} \left\{ 2i|f({\bf p})|^2
\partial_p{\rm Re}\,{\cal F}({\bf p}) \right. \nonumber \\
&& \left. + \left[1 -
2\,{\rm Im}\,{\cal F}({\bf p})\right]f^*({\bf p})
\stackrel{\leftrightarrow}{\partial}_p f({\bf p}) \right\}
\end{eqnarray}
to first order in $e^2$. (Recall that the forward-scattering
amplitude ${\cal F}({\bf p})$ is of order $e^2$.)
We can now add the two
contributions $\langle z\rangle_{\rm em}$ and
$\langle z\rangle_{\rm for}$ together to find
the position expectation value:
\begin{eqnarray}\label{zsum}
\langle z \rangle & = &
\frac{i\hbar}{2} \int\frac{d^3{\bf p}}{(2\pi\hbar)^3} f^*({\bf p})
\stackrel{\leftrightarrow}{\partial}_p f({\bf p}) \nonumber \\
 &&  - \hbar
\int\frac{d^3{\bf p}}{(2\pi\hbar)^3}|f({\bf p})|^2
\partial_p{\rm Re}\,{\cal F}({\bf p})\nonumber \\
 && - \frac{i}{2}\int \frac{d^3{\bf p}}{(2\pi\hbar)^3} |f({\bf p})|^2
\nonumber \\
&& \ \ \ \times
\int \frac{d^3{\bf k}}{2k(2\pi)^3} {\cal A}^{\mu *}({\bf p},{\bf
k}) \stackrel{\leftrightarrow}{\partial}_p {\cal A}_\mu({\bf
p},{\bf k}) \nonumber \\
 && -\frac{i}{2} \int \frac{d^3{\bf p}}{(2\pi\hbar)^3} f^*({\bf
p})\stackrel{\leftrightarrow}{\partial}_p f({\bf p}) \nonumber \\
&& \ \ \ \times \int\frac{d^3{\bf k}}{2k(2\pi)^3} {\cal A}^{\mu
*}({\bf p},{\bf k}){\cal A}_\mu({\bf p},{\bf k}) \nonumber \\
\nonumber \\
&& \ \ \ \times \frac{P_0}{p_0}\frac{dp}{dP}
\left( \frac{dp}{dP} - 1\right)\,.
\label{long}
\end{eqnarray}
The unitarity relation (\ref{unitary}) has been used to eliminate
$2\,{\rm Im}{\cal F}({\bf p})$.
Each term in Eq.~(\ref{long})
can be interpreted as follows. The first term is the
position expectation value $z_0$
for the non-radiating
particle. The second term is the contribution from what can be
regarded as the one-loop quantum correction to the
potential.  Therefore, we identify the sum of the third and forth
terms as the position shift to be compared with the classical
position shift $\delta z_{\rm class}$ given by
Eq.~(\ref{classicalone}).  Using Eq.~(\ref{nonradshift}) and
the assumption that the function $f({\bf p})$ is sharply
peaked about a momentum in the positive $z$-direction with the width
of order $\hbar$, we find these two terms in the $\hbar\to 0$ limit
as
\begin{equation}\label{quantumone}
\delta z_{q} = \delta z_{q1} + \delta z_{q2}\,,
\end{equation}
where
\begin{eqnarray}
\delta z_{q1} & = & - \frac{i}{2}
\int \frac{d^3{\bf k}}{2k(2\pi)^3} \nonumber \\
&& \times {\cal A}^{\mu *}({\bf p},{\bf
k}) \stackrel{\leftrightarrow}{\partial}_p {\cal A}_\mu({\bf
p},{\bf k})\,,\label{deltaq1}\\
\delta z_{q2} & = & - \frac{z_0}{\hbar}
\int\frac{d^3{\bf k}}{2k(2\pi)^3}\nonumber \\
&& \ \ \ \times {\cal A}^{\mu *}
({\bf p},{\bf k}){\cal A}_\mu({\bf p},{\bf k})
\left( \frac{dp}{dP} - 1\right)\,, \label{q2q2}
\end{eqnarray}
where the momentum at which the function $f({\bf p})$ is peaked
is now denoted simply by ${\bf p}$.
We have used the fact that
$dp/dP - 1$ is of order $\hbar$ to drop the factor
$(P_0/p_0)(dp/dP)$ in $\delta z_{q2}$.  We shall demonstrate that
the quantum position shift $\delta z_q$ in the $\hbar \to 0$ limit
is identical with the classical counterpart $\delta z_{\rm class}$
by showing $\delta z_{q1} = \delta z_{\rm LD}$ and
$\delta z_{q2} = \delta z_{\rm extra}$.

Let us first examine the latter equality.  To this end we need to find
an expression for $dp/dP$ in terms of $p$ and $k$.
The energy conservation equation $p_0 - P_0 = \hbar k$ gives a
one-to-one relation between $p$ and $P$ for a given $k$
after letting ${\bf P}_\perp^2 = {\bf p}_\perp^2 = 0$
because these are
of order $\hbar^2$. Then we find
\begin{equation}
\frac{dp}{dP} = 1 - \frac{m^2}{p^2 p_0}\hbar k\,.\label{dpdp}
\end{equation}
By using this formula in Eq.~(\ref{q2q2}) we obtain
\begin{equation}
\delta z_{q2}  =
- \frac{m^2z_0}{p^2 p_0}{\cal E}_{\rm em}\,,  \label{deltazq2}
\end{equation}
where
\begin{equation}\label{emprob}
{\cal E}_{\rm em} \equiv
- \int\frac{d^3{\bf k}}{2k(2\pi)^3} k {\cal A}^{\mu *}({\bf
p},{\bf k}){\cal A}_\mu({\bf p},{\bf k})
\end{equation}
is the expectation value of the energy emitted as radiation.
By comparing Eq.~(\ref{deltazq2})
with Eq.~(\ref{extraaa}), it can be seen that the equality
$\delta z_{q2} = \delta z_{\rm extra}$ will hold if
\begin{equation}\label{larmoreq}
{\cal E}_{\rm em}
=\frac{2\alpha_c}{3} \int_{-\infty}^0 (\gamma^3 \ddot{z})^2\,dt\,,
\end{equation}
which is identical to the relativistic generalization of the classical
Larmor formula.
To show this equality (and also $\delta z_{q1} = \delta z_{\rm LD}$)
we need to find the $\hbar \to 0$ limit of the one-photon-emission
amplitude ${\cal A}_\mu({\bf p},{\bf k})$ to which we now turn.

\section{Emission amplitude}\label{ampcalc}

The one-photon-emission part of the evolution
of the state was represented earlier in Eq.~(\ref{trans}) by
\begin{equation}
A^\dagger({\bf p})|0\rangle \to \cdots + \frac{i}{\hbar}
\int\frac{d^3{\bf k}}{2k(2\pi)^3} {\cal A}^{\mu}({\bf p},{\bf
k})a_\mu^\dagger({\bf k}) A^\dagger({\bf P})|0\rangle\,.
\label{evolA}
\end{equation}
The evolution in
perturbation theory in the interaction picture is generated by the
interaction Lagrangian density as follows:
\begin{equation}\label{evolL}
A^\dagger({\bf p})|0\rangle \to \cdots + \frac{i}{\hbar}\int d^4 x
{\cal L}_I(x)A^\dagger({\bf p})|0\rangle\,.
\end{equation}
We can take the inner product of the two expressions
(\ref{evolA}) and (\ref{evolL})
with the state $\langle 0|a_\nu({\bf k})A({\bf p'})$ and equate
them in order to find the amplitude ${\cal A}^{\mu}({\bf p},{\bf k})$.
This procedure leads to
\begin{eqnarray}
&& - 2\hbar p_0'(2\pi\hbar)^3
\delta({\bf p'}-{\bf P}){\cal
A}_\nu({\bf p},{\bf k}) \nonumber \\
&& = \int d^4x\langle 0|a_\nu({\bf k})A({\bf
p'}){\cal L}_I(x)A^\dagger({\bf p})|0\rangle\,.
\end{eqnarray}
Integrating over the variable
${\bf p'}$, with the appropriate measure,
and rearranging, we find that ${\cal A}_\mu({\bf p},{\bf k})$
is given by
\begin{eqnarray}
{\cal A}_\mu ({\bf p},{\bf k}) & = &
-\frac{1}{\hbar}\int \frac{d^3{\bf p'}}{2p_0'(2\pi\hbar)^3} \nonumber \\
&& \ \ \ \times \int
d^4x\langle 0|a_\mu({\bf k})A({\bf p'}){\cal L}_I(x)A^\dagger({\bf
p})|0\rangle\,.\nonumber \\
\label{Amu1}
\end{eqnarray}
The ${\cal L}_I(x)$ is obtained
from the Lagrangian density (\ref{firstLag}) as follows:
\begin{equation}\label{lint}
{\cal L}_I(x) = -\frac{ie}{\hbar}A_\mu :\left[ \varphi^\dagger D^\mu
\varphi - (D^\mu\varphi)^\dagger \varphi\right]:\,.
\end{equation}
Recall that $D_\mu = \partial_\mu + iV_\mu$, where
$V_\mu = \delta_{\mu 0}V(z)$.
By substituting Eq.~(\ref{lint}) in Eq.~(\ref{Amu1}) and using the
commutation relations (\ref{commu1}) and (\ref{commu2}) we find
\begin{eqnarray}
 {\cal A}_\mu ({\bf p},{\bf k}) &=& -ie\hbar \int \frac{d^3{\bf
p'}}{2p_0'(2\pi\hbar)^3} \int d^4x\,  e^{i k\cdot x}\nonumber \\
&\times &
 \left[ \Phi^*_{\bf p'}(x) D_\mu \Phi_{\bf p}(x) - \left(D_\mu
\Phi_{{\bf p}'}(x)\right)^\dagger \Phi_{\bf p}(x)
\right]\,, \nonumber \\
\end{eqnarray}
where the mode functions $\Phi_{\bf p}(x)$ are given by
Eq.~(\ref{LargePhi}).
As we mentioned before, we assume that the WKB approximation
(\ref{phipp}) is valid for the initial mode function.
Since the momentum $\hbar{\bf k}$ of the
emitted photon is of order $\hbar$, the WKB approximation can also
be used for the mode functions for the
final state.

Since
the transverse momentum ${\bf p}_\perp$ is assumed to be of order
$\hbar$, the final transverse
momentum ${\bf p}_\perp - \hbar {\bf
k}_\perp$ is also of order $\hbar$.
This means that we do not need to consider the $x$- and
$y$-components
${\cal A}_\mu({\bf p},{\bf k})$ because they are smaller by a factor
of $\hbar$ compared
to the $t$- and $z$-components. Using
$D_0 = \partial_t +i V(z)/\hbar$ we obtain
\begin{eqnarray}
{\cal A}_t({\bf p},{\bf k}) & = & - e\int\frac{d^3{\bf
p}'}{2p_0'(2\pi\hbar)^3} \int d^4x \nonumber \\ &\times &
\phi_{p'}^*(z)\phi_p(z) \left[p_0 + p_0'-2V(z)\right] \nonumber \\
&\times & e^{ik\cdot x} e^{-i\left[(p_0'-p_0)t - ({\bf p}'_\perp
 - {\bf p}_\perp)\cdot {\bf x}_\perp \right]/\hbar}\,.
\end{eqnarray}
Integration with respect to $t$, $x$ and $y$ gives
\begin{eqnarray}
{\cal A}_t({\bf p},{\bf k})
& = &
-e\int\frac{dp'}{2p_0'} \int dz \nonumber \\ &&\times
\left[p_0 + p_0'-2V(z)\right]\phi_{p'}^*(z)\phi_p(z) e^{-ik^z z}
\nonumber \\ &&\times \delta(p_0'+\hbar k - p_0)\,.
\end{eqnarray}
We convert the $p'$-integration to
$p'_0$-integration by noting that $dp'/p_0' = dp_0'/p'$ and find
\begin{equation}\label{Atinter}
{\cal A}_t({\bf p},{\bf k})
= -e\int dz\,\frac{P_0+p_0-V(z)}{2P}
\phi_P^*(z)\phi_p(z)e^{-ik^z z}\,,
\end{equation}
where $P_0 = p_0 - \hbar k$ with $P\equiv \sqrt{P_0^2 - m^2}$ as
before.
We let $P_0 =p_0$ and $P=p$ in Eq.~(\ref{Atinter})
because the differences $P_0 - p_0$ and $P-p$ are
of order $\hbar$. Thus, we obtain
\begin{equation}
{\cal A}_t({\bf p},{\bf k})
= -e\int dz\,
\frac{p_0-V(z)}{p}\phi_{P}^*(z)\phi_p(z)e^{-ik^z z} \,.
\label{At}
\end{equation}
Proceeding similarly for
the $z$-component ${\cal A}_z({\bf p},{\bf k})$
we have
\begin{eqnarray}
{\cal A}_z({\bf p},{\bf k}) &=& -\frac{ie\hbar}{2p}\int dz
\nonumber \\
&&\times \left[ \phi_{P}^*(z)\frac{d\phi_p(z)}{dz} -
\frac{d\phi_{P}^*(z)}{dz}\phi_p(z)\right]e^{-ik^z z}\,. \nonumber \\
\label{Az}
\end{eqnarray}
To simplify the expression for the emission amplitude ${\cal A}_\mu$
further, we need to use the explicit form of $\phi_p(z)$ given
by Eq.~(\ref{phipp}).  We note first
\begin{eqnarray}
\phi_{P}^*(z)\phi_p(z) & = &
\sqrt{\frac{Pp}{\kappa_{P}(z)\kappa_p(z)}} \nonumber \\
&& \times \exp\left\{\frac{i}{\hbar}\int_0^z\left[\kappa_{p}(\eta) -
\kappa_{P}(\eta)\right] d\eta\right\}\,.\nonumber \\
\label{phipro}
\end{eqnarray}
Now, to lowest order in $\hbar$ we have
\begin{equation}
\kappa_p(z) - \kappa_{P}(z) =  \frac{\partial
\kappa_p(z)}{\partial p_0}(p_0-P_0) =
\frac{p_0-V(z)}{\kappa_p(z)}\hbar k\,.
\end{equation}
Note that $\kappa_p(z)$ and $p_0 - V(z)$ are the $z$- and
$t$-components of $m\,dx^\mu/d\tau$, respectively, of the classical
particle with final momentum $p$ and vanishing transverse momentum.
Hence, $v_p(z)\equiv \kappa_p(z)/[p_0-V(z)]$ is the velocity of
this classical particle.  Thus,
\begin{equation}
\kappa_p(z) - \kappa_P(z) =
\frac{\hbar k}{v_p(z)}\,.
\end{equation}
By substituting this formula in Eq.~(\ref{phipro}) and letting
$P=p$ and $\kappa_P(z)=\kappa_p(z)$, because the differences
$P-p$ and $\kappa_P(z)-\kappa_p(z)$ are of order $\hbar$, we find
\begin{eqnarray}
\phi_{p'}^*(z)\phi_p(z) &=&
\frac{p}{\kappa_p(z)}\exp\left[
ik\int_0^z\frac{dz}{v_p(z)} \right] \nonumber \\
& = & \frac{p}{\kappa_p(z)}e^{ik t}\,, \label{phiproduct}
\end{eqnarray}
where the time $t$ is defined by $dz/dt = v_p(z)$ with
the condition $t=0$ at $z=0$. Hence
\begin{eqnarray}
 {\cal A}_t({\bf p},{\bf k}) &=& -e\int_{-\infty}^{+\infty}
\frac{dz}{v_p(z)}\, e^{ikt - ik^z z} \nonumber \\
&=& - e \int_{-\infty}^{+\infty}dt\, e^{ikt - ik^z z}\,.
\end{eqnarray}
It is convenient to
define the variable $\xi$ by
\begin{eqnarray}
\xi & \equiv & t - z\cos\theta\,,\\
\cos\theta & \equiv & k^z/k\,.
\end{eqnarray}
Then
\begin{equation}\label{tcomp}
 {\cal A}_t({\bf p},{\bf k}) = - e\int_{-\infty}^{+\infty}
d\xi\, \frac{dt}{d\xi}\, e^{ik\xi}\,.
\end{equation}
Next we turn our attention to the $z$-component given by
Eq.~(\ref{Az}),  We note that, to
lowest order in $\hbar$, we can let
\begin{equation}
\frac{d\phi_p(z)}{dz} = i\frac{\kappa_p(z)}{\hbar}\phi_p(z)\,.
\end{equation}
Hence
\begin{eqnarray}
{\cal A}_z({\bf p},{\bf k}) & = & e\int_{-\infty}^{+\infty} dz
\frac{\kappa_{p}+\kappa_{P}}{2p}\phi_{P}^*(z)\phi_p(z)e^{-ik^zz}
\nonumber \\ & = & e\int_{-\infty}^{+\infty} dz\, e^{ikt-ik^z z}
\nonumber \\
& = &
e\int_{-\infty}^{+\infty}d\xi\,\frac{dz}{d\xi}\,e^{ik\xi}\,,
\end{eqnarray}
where we have let $\kappa_P(z)=\kappa_{p}(z)$ as before and used
Eq.~(\ref{phiproduct}).
This formula and Eq.~(\ref{tcomp}) can be combined as
\begin{equation}\label{emisamp}
 {\cal A}^\mu({\bf p},{\bf k}) = -e\int_{-\infty}^{+\infty}
d\xi\, \frac{dx^\mu}{d\xi}\,e^{ik\xi}\,,
\end{equation}
where $x^\mu$ is the classical trajectory with final momentum $p$
which passes through
$(t,z)=(0,0)$.  (The minus sign for the $z$-component is due to
the raised index.)  This emission amplitude is identical with that for
a classical point charge passing through $(0,0)$~\cite{HM}.

The expression (\ref{emisamp}) is ill-defined since $dx^\mu/d\xi$
remains finite as $\xi\to \pm \infty$.  Therefore,
we introduce a smooth cut-off function $\chi(\xi)$ which takes the
value one
while the acceleration is non-zero with the property
$\lim_{\xi\to\pm\infty}\chi(\xi) = 0$.
Thus,
\begin{equation}\label{cut-off}
{\cal A}^\mu({\bf p},{\bf k}) = -e \int_{-\infty}^{+\infty}
d\xi\, \frac{dx^\mu}{d\xi}\chi(\xi)\,e^{ik\xi}\,.
\end{equation}
In the end we take the limit
$\chi(\xi) \to 1$ at any given $\xi$ in such a way that
$\int_{-\infty}^{+\infty} d\xi\,[\chi'(\xi)]^2 \to 0$.

\section{Derivation of the Larmor formula}\label{larmorsec}

In this section we derive the relativistic generalization of the Larmor
formula for one-dimensional motion, Eq.~(\ref{larmoreq}), thus completing
the demonstration of the equality $\delta z_{q2} = \delta z_{\rm extra}$.
It is convenient to use the following form of the emission amplitude
obtained by integration by parts:
\begin{equation}\label{cut-off2}
{\cal A}^\mu({\bf p},{\bf k}) =
-\frac{ie}{k}\int_{-\infty}^{+\infty}d\xi\, \left[ \frac{d^2
x^\mu}{d\xi^2} +
\frac{dx^\mu}{d\xi}\chi'(\xi)\right]e^{ik\xi}\,,
\end{equation}
where we have used the condition that $\chi(\xi)=1$ if
$d^2x^\mu/d\xi^2 \neq 0$.
By substituting this formula in Eq.~(\ref{emprob}) we obtain
\begin{eqnarray}
{\cal E}_{\rm em}  & = & - e^2\int d\Omega \int_0^\infty
\frac{dk}{16\pi^3} \int_{-\infty}^{+\infty}
d\xi'\int_{-\infty}^{+\infty}d\xi \nonumber \\ && \times
\left[ \frac{d^2 x^\mu}{d{\xi'}^2} +
\frac{dx^\mu}{d\xi'}\chi'(\xi')\right] \nonumber \\
&& \times
\left[ \frac{d^2 x_\mu}{d\xi^2} +
\frac{dx_\mu}{d\xi}\chi'(\xi)\right]e^{ik(\xi-\xi')}\,,
\end{eqnarray}
where $d\Omega$ is the solid angle in the ${\bf k}$-space.
We extend the integration range for $k$ from $[0,+\infty)$ to
$(-\infty,+\infty)$ and divide by two.  Then using
$\int_{-\infty}^{+\infty} e^{ik(\xi-\xi')}\,dk
= 2\pi\delta(\xi-\xi')$, we find
\begin{eqnarray}
{\cal E}_{\rm em}  & = & - \frac{e^2}{16\pi^2}
\int d\Omega\int_{-\infty}^{+\infty}d\xi \nonumber \\ &&
\times
\left\{ \frac{d^2x_\mu}{d\xi^2}\frac{d^2 x^\mu}{d\xi^2} +
\frac{dx_\mu}{d\xi}\frac{dx^\mu}{d\xi}
\left[\chi'(\xi)\right]^2\right\}\,.
\end{eqnarray}
The second term tends to zero in the limit $\chi(\xi)\to 1$ due to the
requirement $\int_{-\infty}^{+\infty}[\chi'(\xi)]^2d\xi \to 0$.  Hence,
we have in this limit
\begin{equation}\label{radia}
{\cal E}_{\rm em}  =
 - \frac{\alpha_c}{4\pi}
\int_{-\infty}^{+\infty}d\xi
\int d\Omega \,\frac{d^2x_\mu}{d\xi^2}\frac{d^2 x^\mu}{d\xi^2}\,,
\end{equation}
where $\alpha_c \equiv e^2/4\pi$ as before.
Now, one can readily show that
\begin{equation}
  \frac{d^2x^{\mu}}{d\xi^2} =
\left(\frac{dt}{d\xi}\right)^3\left[\frac{d\xi}{dt}\frac{d^2
  x^{\mu}}{dt^2}-
  \frac{d^2\xi}{dt^2}\frac{dx^{\mu}}{dt}\right]\,.
\end{equation}
By substituting $d\xi/dt = 1 - \dot{z}\cos\theta$ we find
\begin{eqnarray}
\frac{d^2z}{d\xi^2} &=&
\frac{\ddot{z}}{(1-\dot{z}\cos\theta)^3}\,, \label{d2zdxi2}\\
\frac{d^2 t}{d\xi^2} & = & \frac{d^2z}{d\xi^2}\,\cos\theta\,,
\label{d2tdxi2}
\end{eqnarray}
and hence
\begin{equation}
\frac{d^2x^\mu}{d\xi^2}\frac{d^2x_\mu}{d\xi^2}
= - \frac{\ddot{z}^2 \sin^2\theta}
{(1-\dot{z}\cos\theta)^5}\,\frac{dt}{d\xi}\,.
\end{equation}
By substituting this formula in Eq.~(\ref{radia}) we obtain
Eq.~(\ref{larmoreq}), thus demonstrating the equality
$\delta z_{q2} = \delta z_{\rm extra}$.

\section{Quantum Position shift}\label{qposshiftcalc}

Our next task is to show that $\delta z_{q1} = \delta z_{\rm LD}$.
This will establish that the classical position shift
$\delta z_{\rm class}$
is equal to the quantum one $\delta z_q$ in the $\hbar \to 0$ limit.

In the product
${\cal A}^{\mu*}({\bf p},{\bf k})
\partial_p {\cal A}_\mu({\bf p},{\bf k})$ in Eq.~(\ref{deltaq1}) we use
Eq.~(\ref{cut-off2}) for ${\cal A}^{\mu*}$ and Eq.~(\ref{cut-off}) for
$\partial_p {\cal A}_\mu$.  Proceeding in a way similar to that led to
Eq.~(\ref{radia}) we find
\begin{eqnarray}
\delta z_{q1}
& = &
- \frac{\alpha_c}{4\pi}\int d\Omega \int d\xi
\left\{ \frac{d^2
x^\mu}{d\xi^2}\frac{\partial\ }{\partial p}
\left(\frac{dx^\mu}{d\xi}\right)\right. \nonumber \\
&& \left. \ \ \ \ \ \ \ + \frac{1}{4}\frac{\partial\ }{\partial p}
\left( \frac{dx^\mu}{d\xi}\frac{dx_\mu}{d\xi}\right)
\frac{d\ }{d\xi}\left[\chi(\xi)\right]^2\right\}\,.
\label{shiftandchi}
\end{eqnarray}
Next we demonstrate that the second term, which still contains the
cut-off function, vanishes when integrated over the solid angle
and $\xi$.   Noting that
\begin{equation}
\frac{dx^\mu}{d\xi}\frac{dx_\mu}{d\xi} =
\left(\frac{d\tau}{d\xi}\right)^2
\frac{dx^\mu}{d\tau}\frac{dx_\mu}{d\tau} =
\left( \frac{d\tau}{d\xi} \right)^2\,,
\end{equation}
we find by integration by parts that the contribution of
this term is proportional to the following integral:
\begin{equation}
I \equiv \int d\Omega \int_{-\infty}^{+\infty} d\xi \frac{d\
}{d\xi} \left\{\frac{\partial\ }{\partial p}
\left(\frac{d\tau}{d\xi}\right)^2\right\}[\chi(\xi)]^2\,.
\end{equation}
Since $\partial/\partial p$ is taken with $\xi$ fixed,
the $\xi$- and $p$-derivatives commute.
Hence this integral is equal to
\begin{eqnarray}
I & = & \frac{\partial\ }{\partial p} \int
d\Omega \int_{-\infty}^{+\infty} d\xi
\frac{d\ }{d\xi}\left(\frac{d\tau}{d\xi}\right)^2\,, \nonumber \\
& = & \frac{\partial\ }{\partial p} \left[ \int d\Omega \left(
\frac{d\tau}{d\xi}\right)^2 \right]_{\xi=-\infty}^{\xi=+\infty}\,,
\end{eqnarray}
where we have used the fact that the
$\xi$-derivative of $(d\tau/d\xi)^2$ is
non-zero only if the acceleration is non-zero and, therefore, only
when the cut-off function $\chi(\xi)$ equals one.  Since
\begin{equation}
\int d\Omega \left(\frac{d\tau}{d\xi}\right)^2 =
\frac{4\pi}{(dt/d\tau)^2 - (dz/d\tau)^2} = 4\pi\,,
\end{equation}
we have $I=0$, and thus the contribution from
the second term in Eq.~(\ref{shiftandchi})
vanishes. The remaining term gives the main contribution to the
position shift due to radiation reaction in QED in the $\hbar\to 0$
limit as
\begin{equation}\label{quantshifteqn}
\delta z_{q1} = -\frac{\alpha_c}{4\pi}\int d\Omega \int d\xi
\frac{d^2x^{\mu}}{d\xi^2}
  \frac{\partial}{\partial p} \left( \frac{dx_{\mu}}{d\xi} \right)\,.
\end{equation}

To compare $\delta z_{q1}$ given by this equation with
$\delta z_{\rm LD}$ given by
Eq.~(\ref{deltaz}) we need to find an expression of
$\delta z_{q1}$ in terms of $t$ rather than $\xi$.   Thus, we need
to eliminate the variable $\xi$ using its definition
$\xi = t - z\cos\theta$.
The two components of the
second derivative $d^2x^\mu/d\xi^2$ are given by Eqs.~(\ref{d2zdxi2})
and (\ref{d2tdxi2}).
Let us consider the second factor of the integrand of
(\ref{quantshifteqn}), i.e. the $p$-derivative of
$dx^\mu/d\xi$. By interchanging
the order of integration we obtain
\begin{equation}\label{pdiffxmu}
\frac{\partial\ }{\partial p}\left(\frac{dx^{\mu}}{d\xi}\right) =
\frac{dt}{d\xi}\frac{d}{dt}\left(\frac{\partial
x^{\mu}}{\partial p}\right)_\xi\,,
\end{equation}
where the subscript $\xi$ indicates that the partial derivative
with respect to $p$ is taken with $\xi$ fixed.
Then by differentiating the equation $t = z\cos\theta + \xi$ with
respect to $p$ with $\xi$ fixed, we find
\begin{equation}\label{chaintp}
  \left(\frac{\partial t}{\partial p}\right)_{\xi}
= \left(\frac{\partial z}{\partial p}\right)_{\xi}\cos\theta\,.
\end{equation}
Combining this equation with Eqs.~(\ref{d2zdxi2}) and
(\ref{d2tdxi2}),
which gives $d^2x^\mu/d\xi^2$, we can write the integrand in
Eq.~(\ref{quantshifteqn}) as
\begin{equation}
\frac{d^2x^{\mu}}{d\xi^2}
  \frac{\partial}{\partial p} \left( \frac{dx_{\mu}}{d\xi} \right)
= - \frac{\ddot{z}}{(1-\dot{z}\cos\theta)^4}
\frac{d}{dt}\left(\frac{\partial z}{\partial p}\right)_\xi
  \sin^2\theta\,.
\end{equation}
We substitute this expression in the quantum position
shift (\ref{quantshifteqn}) and change the integration variable
from $\xi$ to $t$ and integrate by parts.  The result is
\begin{eqnarray}
\delta z_{q1}& = & -\frac{\alpha_c}{4\pi}\int d\Omega
\int^0_{-\infty}  dt\, \frac{d\ }{dt}\left[
\frac{\ddot{z}}{(1-\dot{z}\cos\theta)^3}\right]\nonumber \\
&& \ \ \ \ \ \ \ \ \ \ \times
\left(\frac{\partial z}{\partial p}\right)_\xi \sin^2\theta\,,
\label{qshiftsimple}
\end{eqnarray}
where we have also changed the integration range from
$(-\infty,+\infty)$ to $(-\infty, 0]$ because $\ddot{z}=0$ for
$[0,+\infty)$. Finally, we need to relate
$(\partial z/\partial p)_\xi$ to $(\partial z/\partial p)_t$.
By regarding $z$ as a function of $t$ and $p$ we have
\begin{equation}\label{physchem}
dz = \dot{z}dt + \(\frac{\partial z}{\partial p}\)_t dp\,.
\end{equation}
By substituting $dt = d\xi+\cos\theta\, dz$ in this equation and
rearranging, we obtain
\begin{equation}
dz=\frac{\dot{z}}{1-\dot{z}\cos\theta}d\xi
+\frac{1}{1-\dot{z}\cos\theta} \left(\frac{\partial z}{\partial
p}\right)_t dp\,.
\end{equation}
Thus,
\begin{equation}
\left(\frac{\partial z}{\partial p}\right)_\xi =
\frac{1}{1-\dot{z}\cos\theta} \left(\frac{\partial z}{\partial
p}\right)_t\,.
\end{equation}
Using this expression in the position shift (\ref{qshiftsimple})
we have
\begin{eqnarray}\label{3}
 \delta z_{q1}
&=&-\frac{\alpha_c}{4\pi} \int
d\Omega \int_{-\infty}^0 dt \left(\frac{\partial z}{\partial
p}\right)_t \nonumber \\ && \ \ \ \ \ \times
  \[\frac{d\ddot{z}}{dt}\frac{\sin^2\theta}
{\(1-\dot{z}\cos\theta\)^4} + 3
  \ddot{z}^2\frac{\sin^2\theta\cos\theta}
{\(1-\dot{z}\cos\theta\)^5}\]\,. \nonumber \\
\end{eqnarray}
We can now perform the integration over the solid angle
with the following result:
\begin{eqnarray}
  \delta z_{q1} &=& -\frac{2\alpha_c}{3} \int_{-\infty}^0 dt\,
\left[ \gamma^4\frac{d\ddot{z}}{dt}
+3\gamma^6\ddot{z}^2\dot{z}\right]\left(\frac{\partial z}{\partial
p}\right)_t \nonumber \\
 &=& - \int_{-\infty}^0 dt\,
 \[\frac{2\alpha_c}{3}\gamma\frac{d}{dt}
\(\gamma^3\ddot{z}\)\]\left(\frac{\partial z}{\partial
p}\right)_t\nonumber \\
& = & - \int_{-\infty}^0 dt\,
 F_{\rm LD}\left(\frac{\partial z}{\partial
p}\right)_t\,,
\label{qshiftfinal}
\end{eqnarray}
where we have used the expression of $F_{\rm LD}$ given by
Eq.~(\ref{LDLD}).  This formula will be used in the next section to
show that $\delta z_{q1} = \delta z_{\rm LD}$.
We note here that Eq.~(\ref{qshiftfinal}) is valid for an external
force of any form provided
that the photon emission amplitude is given by
Eq.~(\ref{emisamp}).

\section{Comparison of Classical and Quantum Position Shifts}
\label{comparison}

Comparison between Eqs.~(\ref{qshiftfinal})
and (\ref{deltaz}) shows that the equality
$\delta z_{q1}=\delta z_{\rm LD}$ follows if
\begin{equation}\label{concrete}
\left(\frac{\partial z}{\partial p}\right)_t
= \frac{v_0}{m}\dot{z}(t)\int_0^t \frac{dt'}
{\gamma^3(t')\left[\dot{z}(t')\right]^2}\,dt'\,.
\end{equation}
This equation can be demonstrated as follows.
Since the energy is
conserved, we have
\begin{equation}\label{11}
  \sqrt{m^2\left(dz/d\tau\right)^2 + m^2}+V(z) =
  \sqrt{p^2+m^2}\,,
\end{equation}
and, hence,
\begin{equation}\label{dzdteq}
\frac{dz}{dt} =
\left[1-m^2\left(\sqrt{p^2+m^2}-V(z)\right)^{-2}\right]^{1/2}\,.
\end{equation}
By differentiating both sides with respect to $p$ with $t$ fixed,
and noting that
\begin{eqnarray}
p/\sqrt{p^2 + m^2} & = & v_0\,,\\
\sqrt{p^2 + m^2} - V(z) & = & m\gamma\,,
\end{eqnarray}
we obtain
\begin{equation}\label{gamma3}
\frac{d\ }{dt}\left(\frac{\partial z}{\partial p}\right)_t
= \frac{1}{m\gamma^3\dot{z}}
\left[v_0 - V'(z)\left(\frac{\partial z}{\partial p}\right)_t
\right]\,.
\end{equation}
By substituting the formula
$V'(z) = - m\gamma^3 \ddot{z}$ [see Eq.~(\ref{vdash})] in
Eq.~(\ref{gamma3}) we find
\begin{equation}
\frac{d\ }{dt}
\left[\frac{1}{\dot{z}}\left(\frac{\partial z}{\partial p}\right)_t
\right] = \frac{v_0}{m\gamma^3\dot{z}^2}\,.
\end{equation}
Then by integrating this formula, remembering that $z=0$ at $t=0$ for
all $p$, we arrive at Eq.~(\ref{concrete}).

The derivation of $\delta z_{q1} = \delta z_{\rm LD}$ above uses
an explicit relation, Eq.~(\ref{dzdteq}),
between the the trajectory $z(t)$ and the
momentum $p$.  However, the simple form
for the quantum position shift
$\delta z_{q1}$ given by Eq.~(\ref{qshiftfinal}) suggests that there should
be a derivation which does not rely on the relation (\ref{dzdteq}).
We show next that there is indeed such a derivation which is valid
in a more general setting.

We consider a charged particle moving in the $z$-direction
accelerated by an external force which may depend
on both $z$ and $t$.  Thus, we consider
the following equation of motion:
\begin{equation}\label{cleqnmotion}
  m\frac{d^2z}{d\tau^2} = \[F_{\rm ext}(t,z) + F_{\rm LD}\]
  \frac{dt}{d\tau}\,.
\end{equation}
This equation describes a
classical particle under the influence of
an external force $F_{\rm ext}(t,z)$
and the Lorentz-Dirac force.
The model analyzed so far in this paper is the special case given by
$F_{\text{ext}}(t,z) = -V'(z)$.

\begin{figure}
\begin{center}
\begin{pspicture}(-4,-4)(1.5,1.5)
 \psline{->}(-4,0)(1,0) \psline{->}(0,-4)(0,1)
 \uput[r](0,1){$t$} \uput[d](1,0){$z$}
 \pscurve[linewidth=0.5mm]{*-*}(0,0)(-1.2,-1)(-1.8,-2)(-3,-3)
 \pscurve[linewidth=0.5mm]{*-*}(0,0)(-0.6,-1)(-1,-2)(-2,-3)
 \psline{->}(-3,-3.3)(-2,-3.3)
 \uput[d](-2.5,-3.3){$\Delta z_0 (t)$}
 \psline[linestyle=dotted](-4,-3)(0,-3)
 \psline[linestyle=dotted](-3,-3)(-3,0)
 \uput[u](-3,0){$\bar{z}(t)$}
\end{pspicture}
\end{center}
\caption{The world lines for the solutions $(\bar{z}(t),\bar{P}(t))$
and
$(\bar{z}(t)+\Delta z_0(t), \bar{P}(t)+\Delta P_0(t))$} \label{z0t}
\end{figure}
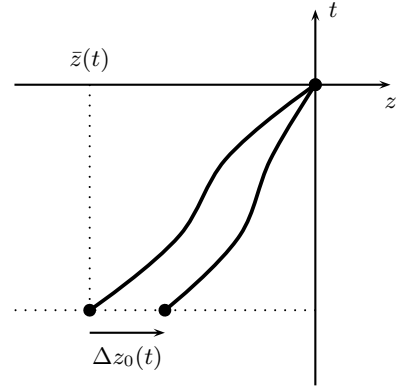

We first consider a solution to Eq.~(\ref{cleqnmotion}) in the
absence of radiation reaction, i.e. with $F_{\rm LD}$ set to $0$,
with the momentum and the position being $p$ and $0$, respectively,
at $t=0$.
We define $P\equiv m\,dz/d\tau$ and
let this solution be denoted
\begin{equation}
  (z,P) = \(\bar{z}(t),\bar{P}(t)\)\,.
\end{equation}
We have $(\bar{z}(0),\bar{P}(0)) = (0,p)$.
Let a linearized solution about $(\bar{z},\bar{P})$,
also with $F_{\rm LD}=0$, be given by
\begin{equation}
  (z,P) = \(\bar{z}+\Delta z,\bar{P}+\Delta P\)\,.
\end{equation}
{}From the equation of motion (with $F_{\rm LD}=0$) the quantities
$\Delta z$ and $\Delta P$ are found to satisfy the equations
\begin{eqnarray}\label{homogenshifts1}
\frac{d}{dt}\Delta z &=& m^{-1}(1-\dot{\bar{z}}^2)^{3/2}\Delta P
\equiv A(t) \Delta P\,,
\\ \frac{d}{dt}\Delta P &=& \frac{\partial F_{\rm ext}}{\partial
z}\Big|_{z=\bar{z}} \Delta z \equiv B(t) \Delta z\,.
\label{homogenshifts2}
\end{eqnarray}
Let $(\Delta z_s(t),\Delta P_s(t))$ be a set of solutions labeled
by $s$ of these equations satisfying $(\Delta z_s(s),\Delta
P_s(s))=(0,1)$. The linearized solutions represent a particle
whose position at $t=0$ coincides with $\bar{z}(0)$ (see
Fig.~\ref{z0t}). The quantity $(\partial z/\partial p)_t$ in
Eq.~(\ref{qshiftfinal}) is the rate of change in the position $z$
at time $t$ as the momentum $P$ at $t=0$ is changed while the
position $z$ at $t=0$ is fixed. Thus,
\begin{equation}\label{qsympldelta}
\(\frac{\partial z}{\partial p}\)_t = \frac{\Delta z_0(t)}{\Delta
P_0(0)} = \Delta z_0(t)\,.
\end{equation}
By substituting this formula in Eq.~(\ref{qshiftfinal}) we obtain
\begin{equation}\label{qshift}
\delta z_{q1} = - \int_{-\infty}^0 dt\,F_{\rm LD}(t)\Delta
z_0(t)\,.
\end{equation}

Next a similar expression will be
derived for $\delta z_{\rm LD}$.
We again let $(\bar{z}(t),\bar{P}(t))$
be a solution of
(\ref{cleqnmotion}) with $F_{\rm LD}= 0$, and now let
$(\bar{z} +\delta z,\bar{P} +\delta P)$
be the approximate solution obtained by
perturbing $(\bar{z},\bar{P})$ to first order in $F_{\rm LD}$
with the initial conditions
$(\delta z, \delta P) = (0,0)$ before the acceleration starts. The
quantities $\delta z$ and $\delta P$
satisfy
\begin{eqnarray}\label{inhomogenshifts1}
\frac{d}{dt}\delta z &=& A(t) \delta P\,,
\\
\frac{d}{dt}\delta P &=& B(t) \delta z + F_{\rm LD}(t)\,,
\label{inhomogenshifts2}
\end{eqnarray}
where $A(t)$ and $B(t)$ are defined by Eqs.~(\ref{homogenshifts1})
and (\ref{homogenshifts2}), respectively.
The quantity $\delta z$ ($\delta P$) is the difference in
position (momentum) between the
hypothetical non-radiating particle and the radiating particle.
Thus, the quantity $\delta z$ at $t=0$ is equal to
$\delta z_{\rm LD}$.

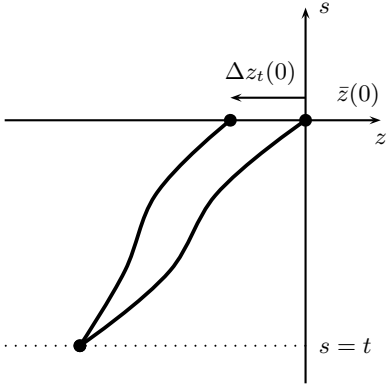
\begin{figure}
\begin{center}
\begin{pspicture}(-4,-4)(2,2)
 \pcline{<-}(-1,0.3)(0,0.3) \uput[u](-0.6,0.3){$\Delta z_t (0)$}
 \psline{->}(-4,0)(1,0) \psline{->}(0,-3.5)(0,1.5)
 \uput[r](0,1.5){$s$} \uput[d](1,0){$z$}
 \uput[u](.7,0){$\bar{z}(0)$}
 \uput[r](0,-3){$s=t$}
 \pscurve[linewidth=0.5mm]{*-*}(0,0)(-1.2,-1)(-1.8,-2)(-3,-3)
 \pscurve[linewidth=0.5mm]{*-*}(-1,0)(-2,-1)(-2.4,-2)(-3,-3)
 \psline[linestyle=dotted](-4,-3)(0,-3)
\end{pspicture}
\end{center}
\caption{The world lines for the solutions
$(\bar{z}(s),\bar{P}(s))$ and
$(\bar{z}(s)+\Delta z_t(s),\bar{P}(s)+\Delta P_t(s))$} \label{zt0}
\end{figure}

The solutions to
Eqs.~(\ref{inhomogenshifts1}) and (\ref{inhomogenshifts2}) with the
initial conditions specified above can be expressed
in terms of another set of solutions
$\(\Delta z_s(t),\Delta P_s(t)\)$ of the non-radiating equations,
Eqs.~(\ref{homogenshifts1}) and (\ref{homogenshifts2}), as
\begin{eqnarray}\label{inhomosoln1}
\delta z(t) & = & \int_{-\infty}^t ds\, F_{\rm LD}(s) \Delta
z_{s}(t)\,,\\ \delta P(t) & = & \int_{-\infty}^t ds\, F_{\rm
LD}(s) \Delta P_{s}(t)\,.\label{inhomosoln2}
\end{eqnarray}
These can readily be shown to satisfy Eqs.~(\ref{inhomogenshifts1})
and (\ref{inhomogenshifts2}) by direct differentiation using
the conditions $\Delta_t z(t)=0$ and $\Delta_t P(t) = 1$.
The classical position shift due to the
Lorentz-Dirac force, at time $t=0$, is therefore given by
\begin{equation}\label{clshift}
\delta z_{\rm LD} = \int_{-\infty}^0 dt\,F_{\rm LD}(t)\Delta
z_{t}(0)\,.
\end{equation}
This expression for the classical position shift $\delta z_{\rm LD}$
is similar to
that for the quantum position shift $\delta z_{q1}$ in
Eq.~(\ref{qshift}), the only difference being that the quantity
$\Delta z_0(t)$ is replaced by $-\Delta z_t(0)$.  In Fig.~\ref{zt0}
we show the world lines corresponding to the solution
$(\bar{z}(s),\bar{P}(s))$ and the approximate solution
$(\bar{z}(s)+\Delta z_t(s),\bar{P}(s)+\Delta P_t(s))$ to the
equations for the hypothetical non-radiation particle motion.

Now, we only need to establish that $\Delta z_0(t)=-\Delta z_t(0)$
to show that the quantum and classical position
shifts (\ref{qshift}) and (\ref{clshift}) are equal.
In fact one can show in general
that $\Delta z_s(t)=-\Delta z_t(s)$ for any values of $s$
as follows. Given a pair of
solutions $(\Delta z^{(1)},\Delta P^{(1)})$ and
$(\Delta z^{(2)},\Delta P^{(2)})$ to
Eqs.~(\ref{homogenshifts1}) and (\ref{homogenshifts2})
we define the symplectic product by
\begin{eqnarray}
&& \langle \Delta z^{(1)},\Delta P^{(1)}|\Delta z^{(2)},\Delta
P^{(2)} \rangle\nonumber \\
&& \equiv \Delta z^{(1)}\Delta P^{(2)}
- \Delta P^{(1)}\Delta z^{(2)}\,.
\label{8}
\end{eqnarray}
It can readily be shown that the symplectic
product is time-independent by differentiating it with respect to $t$
and using Eqs.~(\ref{homogenshifts1}) and (\ref{homogenshifts2}).
With the identification
$(\Delta z^{(1)},\Delta P^{(1)}) = (\Delta z_s,\Delta P_s)$
and $(\Delta z^{(2)},\Delta P^{(2)}) = (\Delta z_t, \Delta P_t)$
we obtain
\begin{eqnarray}
&& \Delta z_s(t)\Delta P_t(t) - \Delta P_s(t) \Delta z_t(t)
\nonumber \\
&& =  \Delta z_s(s)\Delta P_t(s)
- \Delta P_s(s) \Delta z_t(s)\,.
\end{eqnarray}
Since $\Delta z_s(s) = \Delta z_t(t) = 0$ and $\Delta P_s(s)=
\Delta P_t(t)=1$, we have $\Delta z_s(t) = -\Delta z_t(s)$.
Hence the equality $\delta z_{q1} = \delta z_{\rm LD}$ holds.
This equality and the equality
$\delta z_{q2}=\delta z_{\rm extra}$ establish
$\delta z_q = \delta z_{\rm class}$ [see Eqs.~(\ref{classicalone})
and (\ref{quantumone})]. Thus,
the position shift due to QED to order $e^2$
in the limit $\hbar\rightarrow 0$ is equal to the position shift
due to the Lorentz-Dirac force for linear motion if the acceleration
is caused by a static potential.

We have shown
that the classical position shift $\delta z_{\rm LD}$ equals
$\delta z_{q1}$ in  Eq.~(\ref{qshift}) for any external force
which depends on $t$ and $z$.  We have also shown that
Eq.~(\ref{qshift}) is valid {\em provided that the photon
emission amplitude is given by Eq.~(\ref{emisamp})}.  Thus, the
equality $\delta z_{q1} = \delta z_{\rm LD}$ holds if the amplitude
${\cal A}^\mu({\bf p},{\bf k})$ is given by Eq.~(\ref{emisamp}).
We have demonstrated Eq.~(\ref{emisamp})
for the case where the external force
$F_{\rm ext}$ is $t$-independent.  The extension to the case where
the external force depends on both
$t$ and $z$ does not appear to be straightforward.
However, our method can be applied to the case with
an external force that depends only on $t$ because the WKB approximation
is very similar to that for the $t$-independent case as we
demonstrate in the next section.

\section{Time-dependent potential}\label{timeextension}

Let the potential be given by $V_t=0$, $V_z = -V(t)$
in the QED Lagrangian density (\ref{firstLag}).
We show below that the emission amplitude in this
case is also given by Eq.~(\ref{emisamp}).

Since the calculation is quite similar to the time-independent case,
we present only a brief description.
The WKB approximation to the mode functions is given by
\begin{eqnarray}\label{wkbtdep}
\Phi_{\bf p}(t,{\bf x}) & \approx & \tilde{\phi}_{p}(t)
\exp \[\frac{i}{\hbar}\({\bf p}_\perp \cdot {\bf x}_\perp +
p^z z \)\]\,, \\
\tilde{\phi}_{p}(t)
& \equiv &
\sqrt{\frac{p_0}{\sigma_p
(t)}}\exp\[-\frac{i}{\hbar}\int^t_0\sigma_p(\zeta)
 d\zeta\]\,,
\end{eqnarray}
where
\begin{equation}
\sigma_p(t) \equiv \sqrt{\left[p-V(t)\right]^2+m^2}
\end{equation}
if ${\bf p}_\perp=0$. (The mass term $m^2$ needs to be replaced by
$m^2 + {\bf p}_\perp^2$ if ${\bf p}_\perp\neq 0$ as before.)
The function $\sigma_p(t)$ is the
$t$-component of the kinetic energy-momentum $m\,dx^\mu/d\tau$
of the non-radiating classical particle at $t$.
Note that the momentum ${\bf p}$ including the
$z$-component $p^z = p$ is conserved, but the energy is not.
The calculation of the
emission amplitude follows the same pattern as before with
the roles of $t$ and $z$ (and those of $p$ and $p_0$) reversed.

Let ${\cal A}_\mu({\bf p},{\bf k})$ be, again, the emission amplitude
for the process in which the charged particle with momentum ${\bf p}$
emits a photon with momentum $\hbar {\bf k}$ so that the final
momentum of the charged particle is given by
${\bf P}= {\bf p} - \hbar {\bf k}$.  We again assume that the
transverse momenta are of order $\hbar$ and can be neglected.
Then the $z$-component to order $\hbar^0$
[to be compared with Eq.~(\ref{At})] is
\begin{equation}\label{Aztil}
 {\cal A}_z  = e\int dt \,\frac{p-V(t)}{p_0}\tilde{\phi}^*_{P}(t)
\tilde{\phi}_p(t)\,e^{ikt}\,.
\end{equation}
For the $t$-component [to be compared with Eq.~(\ref{Az})]
we obtain
\begin{equation}\label{Attil}
 {\cal A}_t
= -\frac{ie\hbar}{2p_0}\int dt\,\[\tilde{\phi}^*_{P}\partial_t
\tilde{\phi}_p -
 \(\partial_t\tilde{\phi}^*_{P}\)\tilde{\phi}_p\] e^{ikt}\,.
\end{equation}
The product of mode functions in Eq.~(\ref{Aztil}) is
\begin{eqnarray}
\tilde{\phi}^*_{P}(t)\tilde{\phi}_p(t) & = &
\sqrt{\frac{P_0p_0}{\sigma_{P}(t)\sigma_{p}(t)}} \nonumber \\
&&\times
\exp\[-\frac{i}{\hbar}\int^t_0
\(\sigma_p(\zeta)-\sigma_{P}(\zeta)\)d\zeta\]\,. \nonumber \\
\end{eqnarray}
To lowest order in $\hbar$ we can change $P_0$ to $p_0$ and
$\sigma_{P}(t)$ to $\sigma_p(t)$ in the pre-factor
as the differences are of order
$\hbar$.
Recalling the classical momentum conservation equation,
$m\,dz/d\tau + V(t) = p$, and the formula
$m\,dt/d\tau = \sigma_p(t)$, where $z(t)$ is the classical trajectory
with $z(0)=0$, we obtain to lowest order in $\hbar$
\begin{equation}
\sigma_p(t) -\sigma_{P}(t) =
\frac{\partial \sigma_p}{\partial p}(p-P)
 = \frac{dz}{dt}\hbar k^z\,.
\end{equation}
Thus the product of the mode functions in question can be written
\begin{eqnarray}
\tilde{\phi}^*_{P}(t)\tilde{\phi}_p(t) &=&
\frac{p_0}{\sigma_p}\exp\[-\frac{i}{\hbar}\int^t_0 dt\,
\frac{dz}{dt} \hbar
k^z\] \nonumber \\
 &=& \frac{p_0}{\sigma_p}\exp\(-ik^z\, z\)\,.\label{product}
\end{eqnarray}
This gives the $z$-component of the emission amplitude as
\begin{eqnarray}
{\cal A}_z({\bf p},{\bf k})
&=& e\int dt\,\frac{p-V(t)}{\sigma_p(t)}e^{ikt-ik^zz}
\nonumber
\\
 &=& e \int d\xi\, \frac{dz}{d\xi} e^{ik\xi}\,,  \label{Az2}
\end{eqnarray}
where $\xi=t-z\cos\theta$ and $\cos\theta\equiv k^z/k$ as before. To
calculate the $t$-component of the amplitude, note first
that to lowest
order in $\hbar$ we have
\begin{eqnarray}
\hbar\tilde{\phi}_{p'}(t)\partial_t\tilde{\phi}_p(t) & = &
-i\sigma_p(t) \tilde{\phi}^*(t)\tilde{\phi}_p(t)\nonumber \\
& = & -ip_0\exp(-ik^z z)\,,
\end{eqnarray}
where Eq.~(\ref{product}) has been used.
By using this equation in Eq.~(\ref{Attil}) we obtain
\begin{equation}
{\cal A}_t({\bf p},{\bf k})  =  -e\int dt\, e^{ikt - ik^z z}
 = -e \int d\xi\, \frac{dt}{d\xi} e^{ik\xi}\,. \label{At2}
\end{equation}
Eqs.~(\ref{Az2}) and (\ref{At2}) can be written
${\cal A}^\mu({\bf p},{\bf k})= -e \int d\xi\,(dx^\mu/d\xi)e^{ik\xi}$.
As we mentioned before this equality is sufficient to conclude that
$\delta z_{q1} = \delta z_{\rm LD}$.

The equality $\delta z_{q2} = \delta z_{\rm extra}$, which is necessary
to establish the equality of the classical and quantum position shift,
is trivial in this case.
Since the system is translationally invariant, the classical
position shift is invariant under the shift in the position,
$\langle z\rangle = 0 \to z_0$, of
the non-radiation hypothetical particle at $t=0$.  Hence
$\delta z_{\rm extra} = 0$.  Furthermore, since
$P=p-\hbar k^z$, we have $dp/dP =1$ and hence $\delta z_{q2} = 0$
in Eq.~(\ref{q2q2}).  Thus, we have
$\delta z_{q2} = \delta z_{\rm extra}=0$.  Hence,
$\delta z_q = \delta z_{\rm class}$, i.e.
the position shift from
the Lorentz-Dirac force agrees with that in QED in the limit
$\hbar \to 0$ to order $e^2$ for a space-independent potential
as well.

The equality $\delta z_{q1} = \delta z_{\rm LD}$ can be verified
by a more explicit calculation as in the case with a static potential.
Note that the momentum conservation equation reads
\begin{equation}
\frac{d\ }{dt}\left[ m\gamma \dot{z} + V(t)\right]
= F_{\rm LD}\,.
\end{equation}
Hence
\begin{equation}
\delta(m\gamma\dot{z}) =
m\gamma^3\,\frac{d\ }{dt}(\delta z) = \int_{-\infty}^t
F_{\rm LD}(t')\,dt'\,.
\end{equation}
Thus, the classical position shift is
\begin{equation}\label{cshifttdep}
\delta z_{\rm class} = \delta z_{\rm LD}  = -\int^0_{-\infty}
\(\int^t_0\frac{1}{m\gamma^3}dt'\)F_{\rm LD}dt\,,
\end{equation}
where we have interchanged the order of integration.  Thus, what
we need to establish is
\begin{equation}\label{infact}
\(\frac{\partial z}{\partial p}\)_t
 = \int^t_0 \frac{1}{m\gamma^3}\, dt
\end{equation}
for the hypothetical non-radiating particle.
The momentum conservation for this particle in the $z$-direction reads
\begin{equation}
m\frac{dz}{d\tau} + V(t) = p\,.
\end{equation}
Hence, with the condition $z=0$ at $t=0$, we find
\begin{equation}
z = \int^t_0 \left\{ 1+\frac{m^2}{[p-V(t)]^2}\right\}^{-1/2}dt\,.
\end{equation}
By differentiating this expression with respect to $p$ and using
$p-V(t) = m\,dz/d\tau$ and
$\sqrt{[p-V(t)]^2 + m^2} = m\,dt/d\tau$, we indeed obtain
Eq.~(\ref{infact}).

\section{Conclusion}\label{conclusion}
In this paper we showed that the change in position due to the
radiation reaction of a particle, which we call the position
shift, according to the Lorentz-Dirac theory in classical
electrodynamics is reproduced by the $\hbar
\to 0$ limit of QED with a scalar charged particle.  The
calculation was performed to lowest non-trivial order in $e^2$ for
wave packets linearly accelerated for a finite time by either a
purely space-dependent or time-dependent potential. The quantum
wave functions constituting the wave packets were approximated
using the WKB functions for each potential. The agreement between
QED and the Lorentz-Dirac theory was demonstrated by concrete
calculations which gave the position shifts explicitly.  We also
showed by a general argument that this agreement holds as long as
the emission amplitude coincides in the $\hbar\to 0$ limit with
that for a classical point charge.

This work extended the corresponding work in the non-relativistic
approximation~\cite{Higuchi1,Higuchi2}
and provided the details omitted in
Ref.~\cite{HM} as well as the analysis for the time-dependent but
space-independent potential. It will be interesting
to generalize our results to the case with a particle
moving in three space dimensions.
It is also important to estimate the one-loop correction
to the potential.
In the (unrealistic) model in which the acceleration is achieved by a
space-dependent mass term~\cite{Higuchi2}, this
correction is of order $\hbar^{-1}$, thus overwhelming the
contribution from the Lorentz-Dirac force.
If this correction turns out to be of order
$\hbar^0$ in the models studied here, which are more realistic,
then the one-loop correction
will be as important as the Lorentz-Dirac force
in determining the motion of the charged particle.
It will also be interesting to see whether our results can be
justified when more than one photons are emitted: our results, as
they stand, are
logically consistent only if the emission probability
${\cal P}_{\rm em}$ given by Eq.~(\ref{emprobtrue})
is much smaller than
one so that the process is dominated by one-photon emission.
These issues are currently under investigation.

\end{document}